# Significant impact of light-matter strong coupling on chiral nonlinear optical effect

*Daichi Okada [1]\*,    Fumito Araoka[1]\**

[1] *RIKEN Center for Emergent Matter Science (CEMS), 2-1 Hirosawa, Wako, Saitama 351-0198, Japan*

ABSTRACT.

Light-matter strong coupling (LMSC) is intriguing state in which light and matter are coherently hybridized inside cavity. It has been gaining widespread recognition as an excellent way for controlling material properties without any chemical modification. Here we show the LMSC is a powerful state to manipulate and improve a chiral nonlinear optical (NLO) effect through the investigation of second harmonic generation circular dichroism. At the upper polaritonic band in LMSC, in addition to an enhancement of SHG intensity by more than one order of magnitude, the responsivity to the handedness of circular polarized light is largely modified, where sign inversion and increase of dissymmetric factor is achieved. Quarter wave plate rotation analysis reveals that the LMSC clearly influence the coefficients associated with chirality in NLO process and it also contributes to the enhancement of nonlinear magnetic dipole interactions. This study demonstrates that LMSC serves as a novel platform for control of chiral optoelectronics and magneto-optics.

**Introduction**

Chirality is a geometrical description of a structure which cannot be superimposed onto its mirror image. Materials with chirality offers opportunity for the control of electron spin and circular polarized (CP) light, as a result, they show fascinating optical, electronic, and spintronic properties such as spin filtering [1-5], magneto-optical effect [6-8], CP light detection and chiroptical activity [9-14]. Additionally, chirality often induces the symmetry breaking in a structure, leading to an emergence of symmetry-related physical properties, such as ferroelectricity [15-19], even-order nonlinear optical effect (NLO) [20-32] and quantum topological properties [33-37]. Due to the advancements in material synthesis and exploration techniques, we have been able to finely control the physical, chemical, and biological properties in chiral materials, thereby accelerating the development of chiral optoelectronics. However, despite the transformative potential of synthetic technologies in redesigning scientific disciplines, it is important to acknowledge that inherent limits on material's performance are imposed by fundamental physical or device system. Consequently, there is an urgent necessity to incorporate non-synthetic techniques for modulating material properties within the fields of material science, especially in the chiral optoelectronics, to further advance scientific technology. Light-matter strong coupling (LMSC) is a promising candidate as a novel platform, directing the universal way towards non-synthetic manipulation of material properties. LMSC is the hybridized state of light and matter, which can be achieved by coherent energy exchange between an excitonic transition and confined photons inside an optical microcavity. In LMSC, the new quantum state called exciton-polariton is formed and original energy band is split into two anti-crossing energy bands (Rabi-splitting). In past years, it has been reported that LMSC can manipulate and improve various types of materials properties, such as photochemical reactivity [38-41], conductivity [42,43], resonance energy transfer [44,45], nonlinear optical (NLO) effect [46-50], and so on [51-53]. Thus, LMSC has been gaining attention as an emerging technique for control of materials properties. In addition to the above-mentioned case, the exploration of the correlation between LMSC and chirality has been carried out using chiral molecular dyes or chiral plasmonic structures [54-57]. However, despite the potential of LMSC to give a significant impact to various physical properties exhibited by chiral materials, previous studies have predominantly focused on circular dichroism (CD), discussing only the influence of the Rabi splitting on CD spectra. [54-57]

In order to explore the novel potential of LMSC for chiral optical/electronic properties beyond the mere optical activity, we investigated NLO processes manifested as second harmonic generation (SHG) and its circular dichroism (SHG-CD). SHG-CD is one of emergent optical response of the chiral NLO, which is activated differently upon the circularly polarized light stimulation due to the correlation between the chirality and the handedness of the incident circularly polarized light [20-26, 29-32]. Since such coherent discriminability of lights can add an extra degree of freedom of photons participating in the NLO process, it may lead to the innovation in optical/opto-electronic information devices. Here, we demonstrate the chiral NLO response in two-dimensional (2D) chiral organic inorganic hybrid perovskite (OIHP), R/S-(MBACl)$_2$PbI$_4$, confined in a Fabry-Pérot (F-P) type optical microcavity. Due to the formation of the LMSC state, its original absorption peak is split into polaritonic peaks with a large Rabi-splitting energy of 283meV and a coupling strength of 143meV at room temperature. Under LMSC, the SHG intensity is resonantly enhanced one order of magnitude at the upper polaritonic band (UPB) wavelength. In addition, it is found that LMSC in R/S-(MBACl)$_2$PbI$_4$ significantly modulates the responsivity to the CP lights. In the case outside the F-P cavity, the zero-crossing of the dissymmetric factor attributed to the linear CD is confirmed, whereas, in LMSC, not. Instead, the dissymmetric factor increases at the UPB region in accordance with the enhancement of SHG. Carrying out the so-called quarter wave plate (QWP) rotation analysis, it becomes evident that LMSC clearly influences the chirality-related components (chiral coefficients) in the nonlinear susceptibility. This effect causes change in the interference condition of the SH waves originated from the chiral and non-chiral components, leading to the modulation of the dissymmetric response in SHG, that is, SHG-CD. Furthermore, the analysis reveals that LMSC contributes not just to the chiral components, but to the magnetic dipole (MD) interactions. Our finding substantiates that LMSC is indeed a promising approach to optoelectronic materials controlled with the chiral and magneto-optic interactions.

**Results**

In this work, we strongly couple the excitonic transition of the 2D OIHP to the optical confined mode of F-P cavity. The 2D OIHP has a strong excitonic character due to the intrinsic quantum well-like structure, i.e., the perovskite semiconducting layer [PbI$_4$]$^{2-}$ is sandwiched by large band gap organic cations, which is advantageous for the formation of the LMSC state [58-60]. Besides, there are many choices of organic cations for such 2D OIHPs, which provides feasibility to tune their physical properties [58-60]. Here, a chiral 2D OIHP containing R/S-4-Chloro-α-methylbenzylamine R-/S-(MBACl) is used (Figure S1a) [15,27,32]. This chiral OIHP has already been known to exhibit SHG and SHG-CD in thin film state [32]. After dissolving the synthesized single crystals of R-/S-(MBACl)$_2$PbI$_4$ in dimethylformamide (DMF), thin films are fabricated by spin-coating (see details in the materials and methods section). Although the obtained films are polycrystalline, the perovskite layers are stacked and highly oriented on the substrate plane, so that only the (0 0 2$l$) peaks corresponding to the layer stacking length are observed in an X-ray diffractogram (Figure S1b). Both the R- and S-(MBACl)$_2$PbI$_4$ thin films show a strong and sharp absorption peak at about 495nm corresponding to the excitonic transition of the perovskite, and their CD spectra show the Cotton effect and mirror reflections of each other, meaning that the chirality of the incorporated organic cations influences the excitonic property of the perovskite layer (Figure S2). The F-P microcavity structure is schematized in Figure 1a. The R-/S-(MBACl)$_2$PbI$_4$ thin film is sandwiched between a Distributed Bragg reflector (DBR) mirror and a silver mirror (~30nm) with a PMMA protection layer. The DBR mirror of 10.5 pair layers of SiO$_2$/Ta$_2$O$_5$ has been tuned to have a selective reflection band centered at the above excitonic absorption peak of R-/S-(MBACl)$_2$PbI$_4$ (Figure 1b). The induced polaritonic state is investigated by angle-resolved reflectivity. Figure 1c, d shows the angular-dependent reflectance spectra for TM polarization and their contour plot with dispersion extracted from the reflectivity minima, respectively. The reflection spectrum for TE polarization is almost identical. The clear splitting and anti-crossing dispersion behavior suggests the formation of LMSC in the chiral OIHPs inside the F-P cavity at room temperature. The fitting curve obtained by the coupled harmonic oscillator model [61-64] with a negative detuning of -98 meV is also plotted in Figure 1d (black solid line). Using the experimentally-obtained half-widths of excitonic transition peak $\hbar\Gamma_{ex}$= 110meV and the cavity photon peak $\hbar\Gamma_{ex}$= 44 meV (estimated using an empty F-P cavity of only PMMA, Figure S3), we deduce the light matter coupling coefficient $V_A$=145 meV and the Rabi-splitting energy of 283 meV. These values satisfy the necessary condition for LMSC, $V_A^2 > ((\hbar\Gamma_{ex})^2 + (\hbar\Gamma_{cav})^2)/2$. That Rabi-splitting energy and coupling strength are much higher than reported quantum well type GaAs and TMDCs [61-64], that is caused by strong excitonic character of 2D OIHPs. Rabi-splitting in linear CD spectroscopy has also been reported, as well as in spectroscopic photo-absorption, in the case using a F-P cavity or a plasmonic-type resonator [54-57]. The CD spectrum in LMSC (in this case, less DFB layers of 6.5 pair layers of SiO$_2$/Ta$_2$O$_5$ is used) is also modified significantly (Figure S4) – New signal peaks are clearly identified, corresponding to the split excitonic absorption bands in LMSC. The Cotton effect disappears, although the peak intensity doesn't change so much.

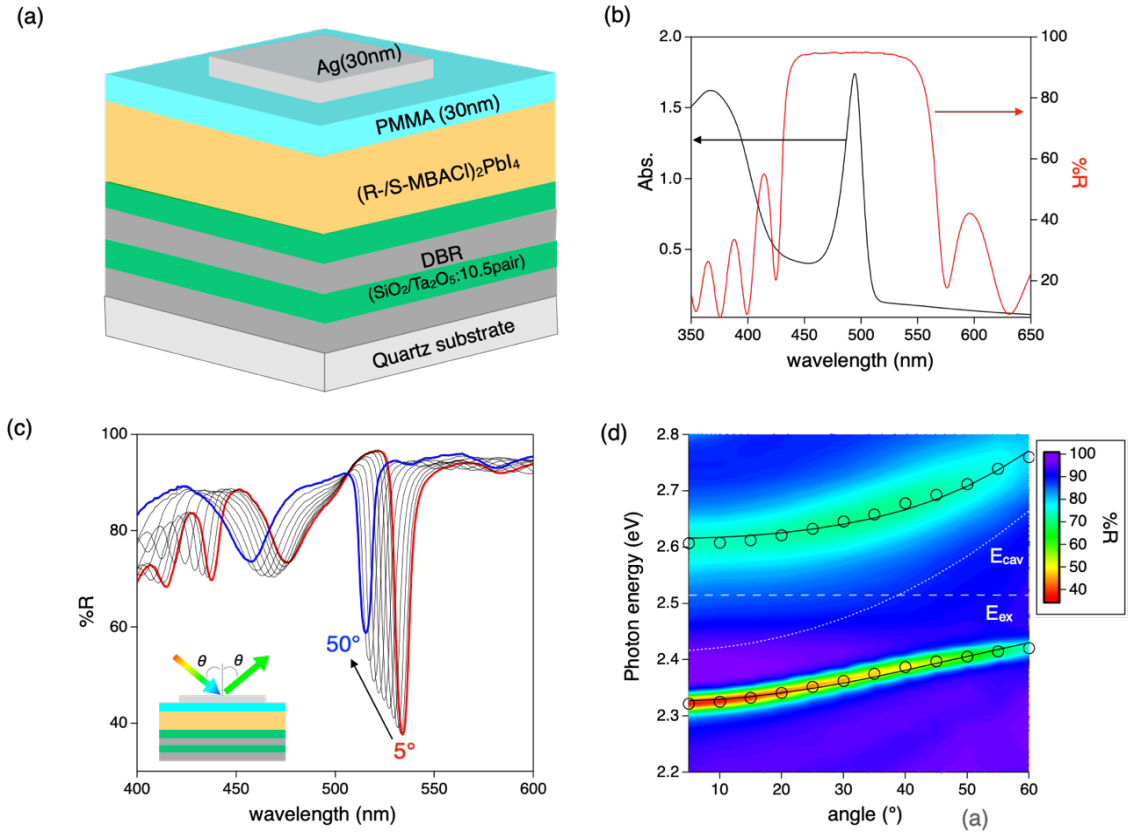

**Figure, 1.** (a) Schematic representation of the F-P cavity. (b) The reflectance spectrum of designed DBR mirrors and photo-absorption spectrum of (R-MBACl)$_2$PbI$_4$ thin film. (c) Angle dependency of reflectance spectrum of optical cavity. (d) Contour plot of the data obtained from Figure 1c. The two splitted reflectance minima is shown as open circle plot. The black solid line is fitting curve using coupled harmonic oscillator model. The dispersion of cavity mode (E$_{cav}$) and Exciton peak top (Ex) is shown as dotted white line.

SHG and SHG-CD are investigated to reveal how LMSC works on the chiral NLO effect. The SHG light is collected in transmission configuration under stimulation of focused fundamental laser pulses (80MHz, 70 fs) irradiated on the substrate surface at 45° incidence. SHG signals are recorded for a fundamental wavelength range from 850 nm to 1050 nm at 10 nm step. For the conventional SHG, the fundamental laser light is *p*-polarized, while for SHG-CD, a quarter waveplate (QWP) is placed before the sample to modulate the fundamental light between the left and right circularly polarizations. The experimental optics is schematized in Figure 2a. First, we examined R-/S-(MBACl)$_2$PbI$_4$ thin films without a F-P cavity structure (hereafter, this condition is termed as "outside-cavity") prepared on a bare quartz substrate as a reference. Both the R- and S-(MBACl)$_2$PbI$_4$ thin films show an up-converted light signal at the half wavelength of the fundamental light. The signal dependence on the fundamental light intensity follows a quadratic function, which proves that this is a result of the two-photon stimulation process, that is, SHG from the samples (Figure S5). The black curve in Figure 2b is a SHG intensity spectrum from the outside-cavity sample upon 120 mW fundamental light stimulation. As the SHG wavelength approaches to the excitonic absorption region, the SHG signal is resonantly pronounced, which peak width is almost identical to the excitonic absorption (Figure S6). The peak top of SHG locates at about 500nm and hence is slightly deviated (red-shifted) from the excitonic peak because self-absorption also happens around the absorption band.

R-/S-(MBACl)$_2$PbI$_4$ also exhibits evident SHG-CD. When using circularly polarized fundamental lights, the SHG intensity largely differs depends on the circular handedness. For quantification, we introduce the dissymmetric factor (g-factor) defined as $g_{SHG-CD}=2(I_L-I_R)/(I_L+I_R)$, where $I_L$ and $I_R$ refer to the SHG intensities measured with left- and right-circularly polarized fundamental lights, respectively. For the fundamental wavelength range from 850 nm to 1050nm, R- and S-(MBACl)$_2$PbI$_4$ give absolute average $g_{SHG-CD}$ values of 0.70 and 0.66, respectively, and draw mirrored spectra of each other. These $g_{SHG-CD}$ values are about three orders of magnitudes higher than the g-factor ($g_{CD}$) in the linear CD (Figure S2c), although the linear CD and SHG-CD have clear correspondences in the sign and the sign inversion wavelengths of the Cotton effect. Note that, a racemic sample, rac-(MBACl)$_2$PbI$_4$, doesn't show SHG-CD at all (Figure S7). Such efficient chiral discrimination is often caused by large contribution of the magnetic dipole (MD) transition in the NLO process [20-24, 32]. It would also be worth noting, our thin films have good optical quality and optical isotropy (but chiral), as confirmed in the fact that SHG is almost unchanged by azimuth rotation (Figure S8). Hence, birefringence is negligible in these R-/S-(MBACl)$_2$PbI$_4$ samples, unlike anisotropic samples presenting the geometric chiral effect [24, 65,66]. In such a way, our present experimental condition can be ideal for pursuit how LMSC influences the chiral NLO effect.

Then, we investigate SHG and SHG-CD in LMSC. The red curve in Figure 2b is the SHG signals for the sample confined in the F-P cavity (hereafter, "inside-cavity"), where SHG is enhanced by one order of magnitude, resonantly at the UPB (see the middle and bottom plates in Figure 2b). The enhancement factor, which is a ratio of the SHG intensity for the outside and inside cavity conditions, is maximized at the peak top wavelength of the UPB. The enhancement of SHG at UPB is confirmed also in other polaritonic detuning conditions (Figure S9), and thus, this is considered as a resonant effect at UPB. In fact, the similar enhancement effect by LMSC has been reported for SHG in a WS$_2$ monolayer [49] and another higher order NLO process, third harmonic generation (THG), in an organic dye [48]. Note that, in contrast to UPB, SHG at the lower polaritonic band (LPB) is more reduced than the outside-cavity condition as also seen in Figure 2b. One plausible explanation could be like this - In the excitation state dynamics in LMSC, the dumping characteristics differ between UPB and LPB. The former proceeds instantaneously through the kinetic process, but the latter undergoes the thermodynamic relaxation [61]. Meanwhile, the nonlinear susceptibility for SHG is generally written by, $\chi^{(2)} \propto \frac{1}{(\omega_0^2-\omega^2-i\Gamma\omega)^2(\omega_0^2-4\omega^2-i2\Gamma\omega)}$, where $\omega$, $\omega_0$, and $\Gamma$ are the fundamental and resonant frequencies, and the dumping factor, respectively. Thus, if the dumping becomes dominant, that is, the case of LPB in the present study, the corresponding resonant SHG peak broadens and lowers in the spectral range. On the other hand, UPB has less dumping so as to work efficiently to enhance SHG in the resonance condition.

LMSC also critically affects SHG-CD at UPB. In the case of the outside-cavity condition, as mentioned, there is a zero-crossing point due to the sign inversion of $g_{SHG-CD}$, accordingly to the Cotton effect of the linear CD spectrum (Figure 2c, S2b, S2c). However, in LMSC, the sign inversion of $g_{SHG-CD}$ doesn't occur, and $|g_{SHG-CD}|$ increases through the UPB. As shown in Figure 2c, in comparison with the outside-cavity condition, the maximum and average $|g_{SHG-CD}|$ values change from 1.57 to 0.95, and from 0.70 to 0.59, respectively, in R-$(MBACl)_2PbI_4$, and from 1.42 to 0.94, and 0.66 to 0.55, respectively, in S-$(MBACl)_2PbI_4$. Thus, the values of $g_{SHG-CD}$ itself tend to be smaller in the inside-cavity condition with LMSC than in the outside-cavity condition without LMSC. However, due to the absence of the zero-crossing, $|g_{SHG-CD}|$ is substantially increased near the sign inversion wavelength compared to the outside-cavity. Such modification of SHG-CD is not induced merely by the contribution of DBR, since the R-/S-$(MBACl)_2PbI_4$, films prepared on the top of DBR without the F-P cavity structure shows the similar behavior to that on a bare quartz substrate (Figure S10). The enhancement of SHG in the inside-cavity condition is often explained by the confinement effect of the electromagnetic waves of the fundamental and/or SHG light [47]. On the other hand, the spectral modification of SHG-CD is considered to be the results of LMSC manipulating the intrinsic chiral-optic process in the NLO effect, because of the concurrency of the aforementioned disappearance of the sign inversion and the accordance of the peak positions in the linear CD and SHG-CD spectra, i.e., the modulation of the linear CD and SHG-CD are obviously linking each other.

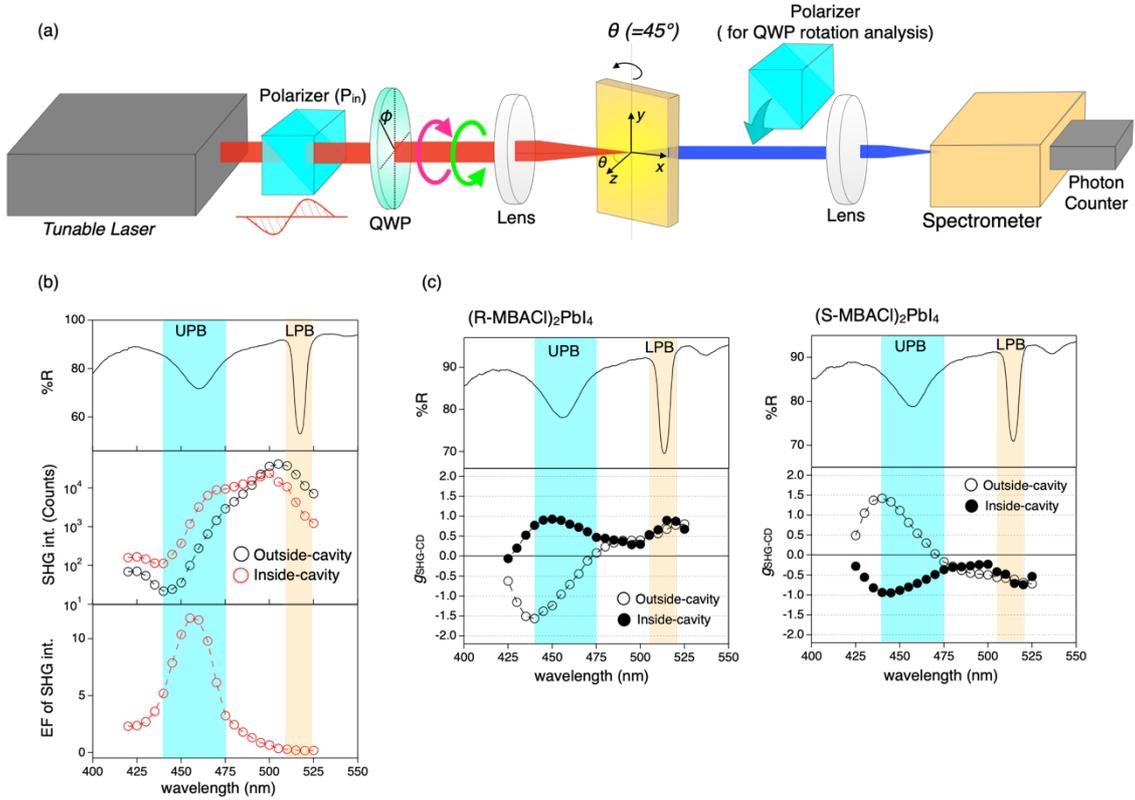

**Figure 2,** (a) Schematic representation of optical set-up for SHG and SHG-CD measurement. (b) SHG intensity spectrum of inside and outside cavity (middle) and enhancement factor by LMSC (SHG intensity inside cavity is divided by that outside-cavity) (bottom). (c) Dissymmetric factor, $g_{SHG\text{-}CD}$, as function of SHG wavelength of (R/S-MBACl)$_2$PbI$_4$ thin film. The top figure in Figure 2b, c shows reflectance spectrum of LMSC at 45° incidence.

In order to gain the deeper understanding of the LMSC effect on SHG-CD, we carried out the so-called QWP rotation (QWR) analysis [20-24], by which the polarization of the fundamental light is continuously changed by rotating the QWP. By doing this analysis, the nonlinear susceptibility tensor components contributing to the anisotropic/chiral response in SHG/SHG-CD are identified and quantitatively evaluated. Also, as already mentioned, the present thin film sample are optically isotropic, making it possible to evade the unpreferred effect of the geometric chirality and to discuss the intrinsic chiral effect in the NLO process. In general, intensity of the transmitted SHG is conveniently described in terms of the *p*- and *s*-polarized components of the fundamental light field as follows, [20-24]

$$I(2\omega) = \left[l(\theta)E_p(\omega)E_p(\omega) + m(\theta)E_s(\omega)E_s(\omega) + n(\theta)E_p(\omega)E_s(\omega)\right]^2 . \quad (1)$$

Where $l(\theta)$, $m(\theta)$, and $n(\theta)$ are symmetry-dependent complex parameters, each of which being expressed as a linear combination of the nonlinear susceptibility tensor components at an incident angle $\theta$ (45° in the present case). For an optically-isotropic chiral film with a space group $C_\infty$, $l(\theta)$, $m(\theta)$ and $n(\theta)$ can be expressed with the linear combinations of certain susceptibility tensor components for *p*- and *s*-polarized SHG. [20-24]

$$l_p = sin\theta[\chi^{eee}_{zzz}sin^2\theta + \chi^{eee}_{zxx}cos^2\theta + 2\chi^{eee}_{xxz}cos^2\theta - \chi^{eem}_{zxy}cos\theta - \chi^{eem}_{zxy}cos\theta + 2\chi^{mee}_{xyz}cos\theta]$$

$$m_p = sin\theta(\chi^{eee}_{zxx} - \chi^{eem}_{zxy}cos\theta + \chi^{eem}_{xyz}cos\theta)$$

$$n_p = sin\theta[2\chi^{eee}_{xyz}cos\theta + \left(\chi^{eem}_{zzz} - \chi^{eem}_{zxx}\right)sin^2\theta + \left(\chi^{eem}_{xzx} - \chi^{eem}_{xxz}\right)cos^2\theta - 2\chi^{mee}_{xxz}]$$

$$l_s = sin\theta[-2\chi^{eee}_{xyz}cos\theta - \chi^{eem}_{xzx} + \chi^{mee}_{zzz}sin^2\theta + \chi^{mee}_{zxx}cos^2\theta + 2\chi^{mee}_{xxz}cos^2\theta]$$

$$m_s = sin\theta(\chi^{eem}_{xxz} + \chi^{mee}_{zxx})$$

$$n_s = sin\theta[2\chi^{eee}_{xxz} - \left(\chi^{eem}_{xzy} + \chi^{eem}_{xyz}\right)cos\theta + 2\chi^{mee}_{xyz}cos\theta] \tag{2}$$

Here, the subscription *p* and *s* represent *p*- and *s*-polarization of SHG, respectively. Thus, in the present analysis, an output polarizer was placed right behind the sample to resolve the SHG output into *p*- and *s*-polarized components (Figure 2a). It is well-known that chirality often induces the MD transition moment in addition to the electric dipole (ED) transition process, so that we consider three possible types of virtual transition schemes as diagramed in Figure S11, that is, ED+ED→ED, ED+MD→ED, and ED+ED→MD transition processes, characterized by specific nonlinear susceptibility tensors denoted as $\chi^{eee}$, $\chi^{eem}$, and $\chi^{mee}$, respectively [20-24]. Since the fundamental light field, $E_p(\omega)$ and $E_s(\omega)$, are functions of the rotation angle of QWP, this analysis allows us to determine the unique values of $l(\theta)$, $m(\theta)$ and $n(\theta)$ (see Supporting S3). Importantly, among these parameters, $l_s$, $m_s$, and $n_p$ are nonvanishing only for chiral symmetries, while $l_p$, $m_p$, and $n_s$ remain in any isotropic systems. Thus, the former are, so to say, "chiral coefficients" in the nonlinear susceptibility, and the latter, "achiral components". Since the degree of the dissymmetric response in SHG-CD is described as $\Delta I_{SHG-CD} \propto imag[(-l + m)n^*]$, one may clearly see that SHG-CD is a manifestation of the phase difference of the waves originated from the chiral and achiral coefficients [20-24] (detailed in Supporting S3). In the case of an isotropic medium, the diagonal nonlinear susceptibility tensor components in *p*-polarized condition are dominate the overall SHG signals. Thus, in SHG-CD as well, anisotropic response to CP lights in *p*-polarized condition is dominant and exhibits almost identical $g_{SHG-CD}$ spectral shapes to that without output polarizer (Figure S12). Therefore, we recorded the *p*-polarized SHG of (S-MBACl)$_2$PbI$_4$ as function of QWR angle in the range of $\lambda_{SHG}$= 425~525 nm. The obtained plots for the wavelengths corresponding to UPB ($\lambda_{SHG}$= 450 nm and 460 nm) and LPB ($\lambda_{SHG}$= 515 nm), and the excitonic transition wavelength ($\lambda_{SHG}$=495 nm) are compared in Figure 3 (see Figure S13 for other wavelengths). The profiles for LPB at 515 nm for the inside- (Figure 3e) and outside- (Figure 3j) cavity conditions are mutually similar. Almost the same trend can be found also in the excitonic region at 495 nm (Figure 3d, 3i). However, the cases for UPB at 450 nm (Figure 3b, 3g) and 460 nm (Figure 3c, 3h) are apparently different, i.e., the profiles for the inside- and outside-cavity conditions in UPB totally distinct from each other. Also, their variation upon varying wavelength is also different - the profiles for the outside-cavity condition are gradually deformed to the more complex shapes as detuned from the excitonic region as seen in Figure 3 and S13, while not in the inside-cavity condition. By comparing with absorption and CD spectra (Figure S2), one may notice that this deformed region corresponds to the gap with low absorption between the two optically resonant conditions peaking at 495nm and 375 nm. These QWR profiles are fitted using Eq. (1), and the relative chiral coefficient, $n_p$, is evaluated as normalized by the achiral coefficient, $l_p$ (Figure 4a, b). In the above gap region in the outside-cavity condition, $n_p$ shows continuous change depending on the wavelength and sign inversion in both the real and imaginary components. Because this region is mixing state of resonant- and non-resonant contributions, as a result, complex phase change occurs in SHG waves. But in the case of the inside-cavity condition, both the real and imaginary components of $n_p$ take almost comparable values with same sign throughout the entire measured wavelength range from 425 to 525 nm. This is likely because that the formation of UPB transitions the non-resonant contribution in the gap between the two optical resonant peaks into a resonant one. These QWR results suggest that the modulation of $g_{SHG-CD}$ at UPB is not a simple sign inversion, but rather the result of the clear impact from LMSC to the chiral NLO process. It should be emphasized that, the $n_p$ value is almost negligible in the case of the racemic (MBACl)$_2$PbI$_4$ sample, while the R-/S-(MBACl)$_2$PbI$_4$ samples exhibit finite $n_p$ values with same magnitude and opposite signs to each other chirality (Figure S14, Table S1). Thus, it is safely concluded that LMSC affect the chiral NLO process. From the results in Figure 4a, we can deduce PD between the chiral and achiral coefficients both in the inside- and outside-cavity conditions (Figure 4b). For comparison, it would be good to define and use a modulation ratio (MR) both for PD and $g_{SHG-CD}$, a ratio of the values for the inside-cavity and outside cavity conditions (Figure 4c). One can see that MR for PD and $g_{SHG-CD}$ well matches each other (Figure 4c), elucidating that the modification of SHG-CD is attributed to the induced PD through LMSC.

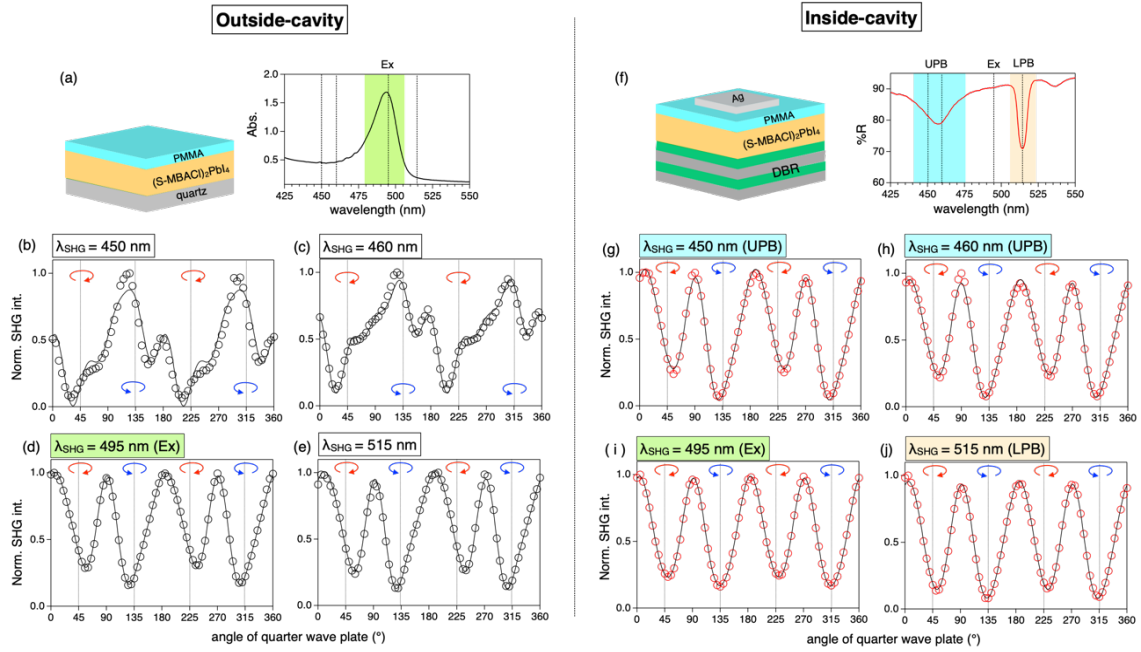

**Figure 3,** (a, f) Schematic of sample inside-(a) and outside-(f) cavity and absorption (a) or reflectance (f) spectrum showing recorded SHG wavelength corresponding LPB (515 nm), bare excitonic transition (495 nm) and UPB (450, 460 nm). (b-e, g-j) The normalized SHG intensity profile as function of QWP rotation angle outside-cavity (b-e) and inside-cavity (g-j). The dot shows experimental result and black solid line shows fitting curve using Eq. (1) transformed as function of QWP rotational angle (detailed in Supporting information S3)

At last, we discuss the origin of such PD. As already expressed in Eq. (2), the coefficients *l, m,* and *n* are the linear combinations of nonlinear susceptibility tensor components both for the ED and MD processes. However, one may notice that in Eq. (2), the terms including $\chi^{eee}$ can be eliminated simply by summing two chiral coefficients $n_p$ and $l_s$, and then there remains only the tensor components for MD related to chirality. Figure 4d shows a plot of an MR of $|n_p + l_s|$, defined as a ratio between the inside- and outside-cavity conditions in the UPB region. The shape of the MR of $|n_p + l_s|$ has a similar trend to the absolute MR of $g_{SHG-CD}$ (see the inset of Figure 4c), both of which peaking near the zero crossing of the $g_{SHG-CD}$ spectrum. Thus, we see in LMSC, the contribution of MD is modified to be more pronounced in the chiral NLO process. In other words, MD plays a vital role in SHG-CD, consequently, in the chiral NLO process, the larger contribution of MD results in the larger dissymmetric response in SHG [20-24, 32].

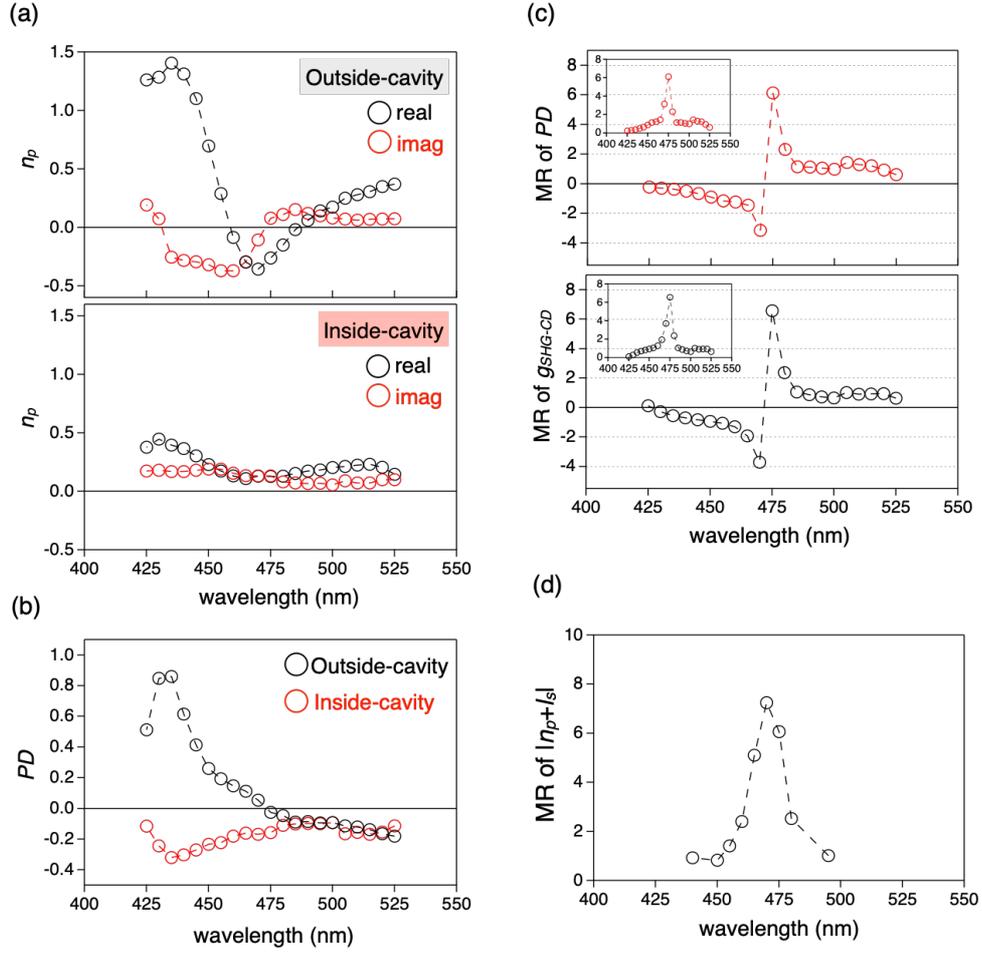

**Figure 4,** (a) The evaluated fitting value of chiral coefficient, $n_p$, of outside and inside cavity with real and imaginary parts. (b) Phase difference (PD) calculated by $imag[(-l+m)n^*]$. (c) Modulation ratio (MR) of PD and $g_{SHG\text{-}CD}$. Inset shows absolute value of each spectrum. (d) MR of $|n_p + l_s|$.

**Discussion**

In summary, our study demonstrates the potential of LMSC as an effective way for manipulating chiral NLO, with a specific focus on investigating the SHG-CD effect. The formation of LMSC has yielded efficiency improvements in both SHG and SHG-CD, in addition to affording control over the sign of the dissymmetric response. These enhancements are particularly manifested at the UPB, this might be caused by the fact both UPB and SHG are obtained by instantaneous processes, not like thermodynamic relaxations (dumping). The QWR analysis provides evidence that LMSC not only clearly influences the chirality-related part of NLO but also exerts an impact on MD interactions, thereby contributing to a modulation of the SHG-CD effect. These findings underscore the efficacy of LMSC as a powerful tool for the non-synthetic manipulation of diverse chiral optoelectronic phenomena, offering a pathway for advancing various scientific fields.

**Materials and Methods**

**Preparation of (R/S-MBACl)$_2$PbI$_4$ single crystals and thin films**

Unless otherwise noted, all reagents and solvents were used as received. (R)- and (S)- 4-Chloro-α-methylbenzylamine (R-/S-MBACl) is purchased from Thermo scientific （Alfa Aesar）. 57% w/w HI aqueous and 50% H$_3$PO$_2$ aqueous solution, PMMA are purchased from TCI. PbO and anhydrous DMF are purchased from Kanto Chemical Industry.

Before preparation of thin films, we synthesized single crystals. For the synthesis of single crystal, 200 mg PbO powder and 200uL R- or S-MBACl were dissolved in 8 mL HI solution with 1mL H$_3$PO$_2$ solution as stabilizer at 140°C until the transparent yellow solution is formed. Then the solution is slowly cooled to room temperature (25°C) for 48 hours. The yellow crystals are formed. The obtained crystals are filtrated and washed by toluene, then it was dried in vacuum for more than 24hours.

For preparation of thin film, the synthesized (R- or S-MBACl)$_2$PbI$_4$ crystals were dissolved in DMF at 200mg/ml concentration, the thin film was obtained by spin-coating 4000rpm for 60s and subsequent annealing at 80°C for 15min on quartz or dielectric mirror coated substrate. To prevent the direct contact to deposited top metal mirror, a 30 nm PMMA film was spin-coated on top of the chiral OIHPs film. Thin film preparation is conducted inside a N$_2$ filled glovebox.

**SHG and SHG-CD measurement**

A Ti:Spphire laser system (Coherent:chameleon) with a pulse width of 70 fs and a repetition rate of 80MHz was used to stimulate the thin film samples. The measurement is conducted by transmission configuration. Laser source is fixed to *p*-polarization by Glan laser prism and sample is irradiated at 45°. In order to measure SHG-CD, achromatic quarter waveplate is placed between sample and polarizer before sample stimulation. SHG is collected by a photon counter through a spectrometer (Andor: Kymera 328i), (Figure 2a). For the QWP rotation analysis, another polarizer is placed before the detector to separate *p*- or *s*-polarized components of SHG.

AUTHOR INFORMATION
**Corresponding Author**
**Daichi Okada**: RIKEN Center for Emergent Matter Science (CEMS),
2-1 Hirosawa, Wako, Saitama 351-0198, Japan
E-mail: daichi.okada @riken.jp

**Fumito Araoka**: RIKEN Center for Emergent Matter Science (CEMS),

2-1 Hirosawa, Wako, Saitama 351-0198, Japan

E-mail: fumito.araoka@riken.jp

**Author Contributions**

D. O and F. A contributed equally this work.

**Funding Sources**

JSPS KAKENHI (JP 23H01942, JP 21K14605, JP21H01801), JST CREST (JPMJCR17N1, JPMJCR23O1), JST SICORP EIG CONCERT-Japan (JPMJSC22C3)

**Notes**

The authors declare no competing financial interest.

ACKNOWLEDGMENT

This work was partly supported by Grant-in-Aid for Scientific research (B) (JP 23H01942, JP21H01801) and Young Scientists (JP21K14605) from Japan Society for the Promotion of Science (JSPS), JST CREST (JPMJCR17N1, JPMJCR23O1), and JST SICORP EIG CONCERT-Japan (JPMJSC22C3).

# Supplementary Materials

## Significant impact of light-matter strong coupling on chiral nonlinear optical effect

Daichi Okada,[1]* Fumito Araoka, [1]*

*Corresponding author. Email: daichi.okada@riken.jp, fumito.araoka@riken.jp

**Table of Contents**

## 1. Material's structure and linear optical response

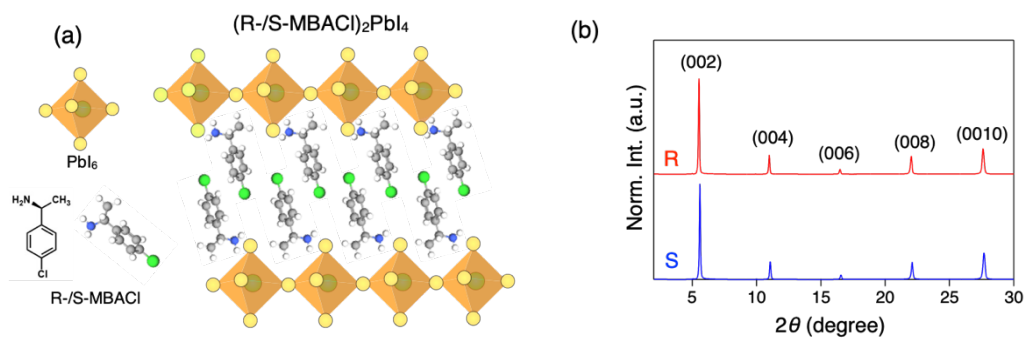

**Figure S1,** (a) Schematic representation of chiral 2D perovskite. (b) XRD pattern of (R/S-MBACl)$_2$PbI$_4$ spin-coated on quartz substrate.

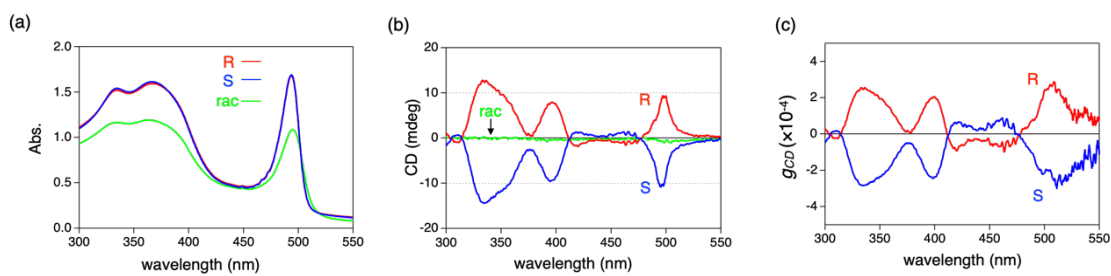

**Figure S2,** (a, b) Photo-absorption (a) and CD (b) spectrum of (R/S/rac-MBACl)$_2$PbI$_4$ thin film. (c) CD spectrum converted to g-value

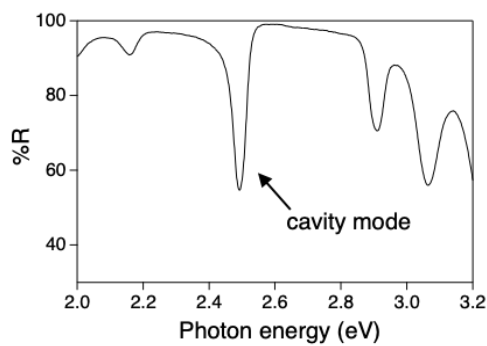

**Figure S3**. Optical cavity mode of empty cavity. The transparent polymer (PMMA) is introduced inside cavity.

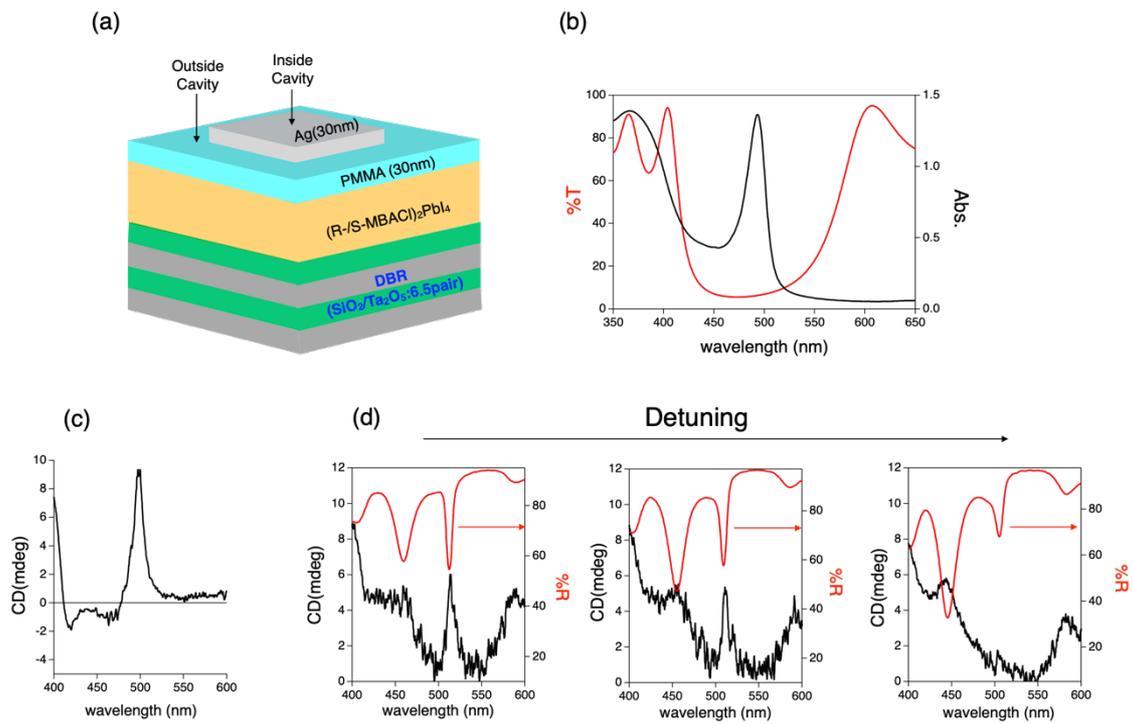

**Figure S4,** (a) Schematic representation of less stacked cavity structure for CD measurement (b) Overlapped spectrum of DBR reflectance and photo-absorption. (c) CD spectrum of (RMBACl)$_2$PbI$_4$ samples. (d) CD spectrum (black) and rabi-splitting spectrum (red) in LMSC state at different detuning conditions. These data are recorded at 5° incidence. The peak with same sign at LPB and UPB wavelength is confirmed in LMSC state, and their peak is shifted according to the shift of rabi-splitting peak by cavity detuning.

## 2. SHG and SHG-CD measurement

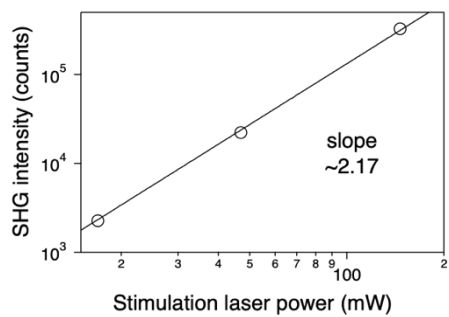

**Figure S5,** Stimulation laser power dependency of SHG at $\lambda_{SHG}$=495nm.

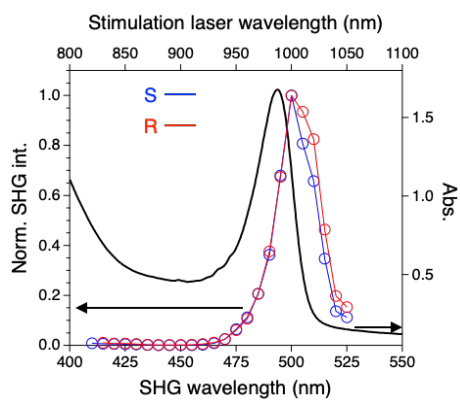

**Figure S6,** Normalized SHG intensity spectrum at 120mW stimulation laser power. It is same plot show in Figure 2b (middle) but not in a log plot.

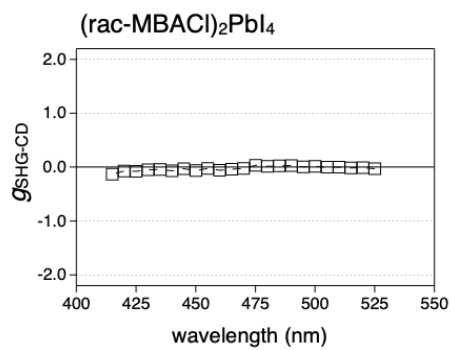

**Figure S7,** SHG-CD performance of (rac-MBACl)$_2$PbI$_4$ thin film

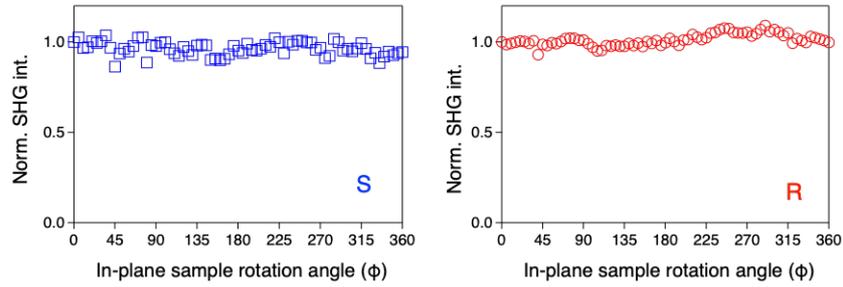

**Figure S8,** SHG as function of in-plane sample rotation angle at $\lambda_{SHG}$=495nm. The SHG intensity is almost independent at any rotational angle, indicating our obtained thin film is optically isotropic.

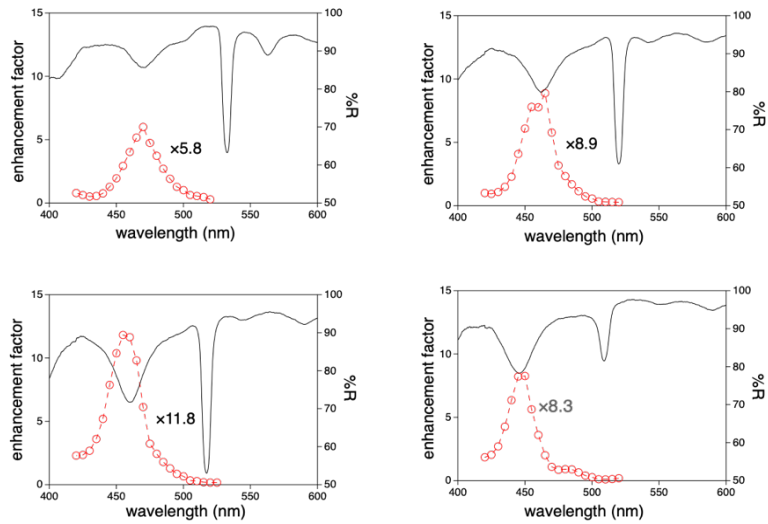

**Figure S9,** The enhancement factor of UP resonance SHG under different detuning conditions. The SHG are most enhanced at the UPB peak top.

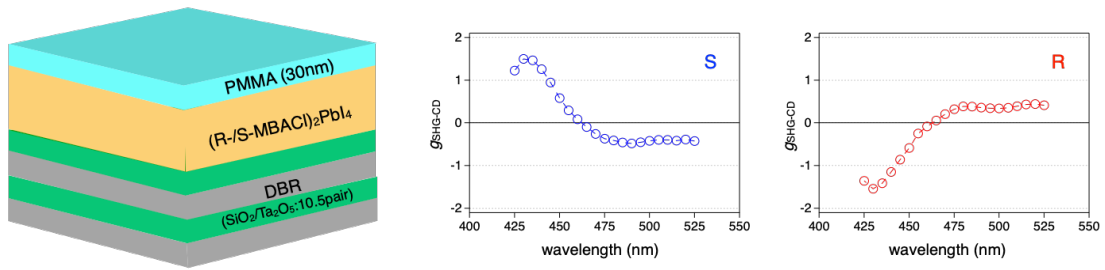

**Figure S10,** $g_{SHG\text{-}CD}$ spectrum of thin film spin-coated on DBR substrate. The obtained data show similar tendency with samples on quartz substrate.

### 3, QWP rotation analysis

**QWP rotation (QWR) analysis**

As already mentioned in main text, the intensity of the SH light generated by a surface can be expressed in the general form as follows,

$$I(2\omega) = [l(\theta)E_p(\omega)E_p(\omega) + m(\theta)E_s(\omega)E_s(\omega) + n(\theta)E_p(\omega)E_s(\omega)]^2 \quad (S1)$$

Since $E_p(\omega)$ and $E_s(\omega)$ are expressed by functions of the rotation angle, $\varphi$, of the QWP, Eq.(S1) is rewritten in terms of the complex fields, $L(\varphi)$, $M(\varphi)$ and $N(\varphi)$, as follows, [20-24].

$$I(2\omega) = [l(\theta)L(\varphi) + m(\theta)M(\varphi) + n(\theta)N(\varphi)]^2 \quad (S2)$$
$$L(\varphi) = E_0^2(\sin^2\varphi + i\cos^2\varphi)^2$$
$$M(\varphi) = E_0^2\sin^2\varphi\cos^2\varphi(1-i)^2$$
$$N(\varphi) = E_0^2(\sin^2\varphi + i\cos^2\varphi)\sin\varphi\cos\varphi(1-i)$$

Thus, QWP rotation analysis evaluate the indexes $l(\theta)$, $m(\theta)$ and $n(\theta)$ ($\theta=45°$ in this study)

**MD interaction in NLO process**

In order to describe the SH light conversion including MD transition, we must take into consideration of the MD interaction in nonlinear polarization (S3) and nonlinear magnetization (S4), and they are expressed as follows,

$$P_i(2\omega) = \chi_{ijk}^{eee}E_j(\omega)E_k(\omega) + \chi_{ijk}^{eem}E_j(\omega)B_k(\omega) \quad (S3)$$
$$M_i(2\omega) = \chi_{ijk}^{mee}E_j(\omega)E_k(\omega) \quad (S4)$$

where E(ω) and B(ω) are the electric field and magnetic induction fields, respectively. The superscript in the susceptibility component associates the respective subscripts with electric dipole (*e*) and magnetic dipole (*m*) interactions. Both nonlinear polarization and magnetization act as origin of SHG [20-24]. Each transition process is schematically shown in Figure S11.

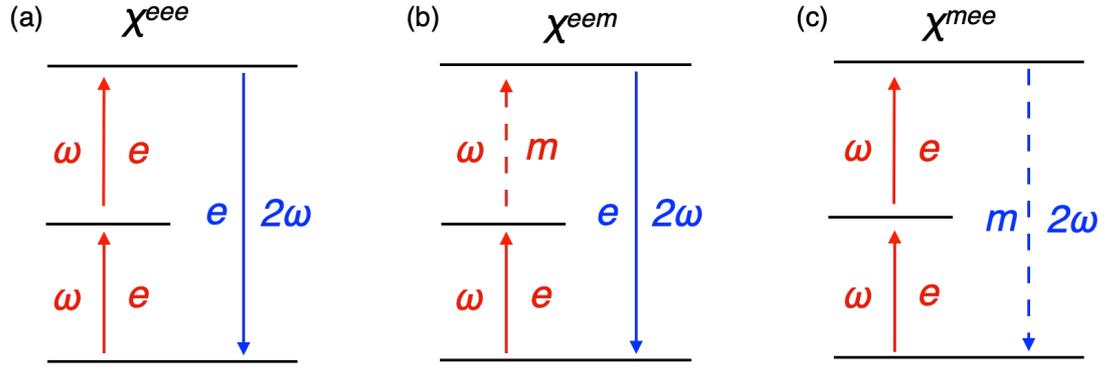

**Figure S11,** (a) Nonlinear polarzation process via the coupling of two EDs (Eq.S3, 1st term) (b) Nonlinear polarization process via the coupling of ED and MD, respectively (Eq.S3, 2nd term) (c) Nonlinear magnetization process via the coupling of two EDs (Eq. S4)

### SHG-CD effect from chiral thin film

The SHG intensity is described as Eq. (S1). The index of $l$, $m$ and $n$ are various linear combination of three types of second order susceptibility component, $\chi_{ijk}^{eee}$, $\chi_{ijk}^{eem}$, $\chi_{ijk}^{mee}$, as shown in Eq.(S3) and (S4), and second order susceptibility component is complex value.

If we consider circular polarized stimulation for which $E_p(\omega) = \pm i E_s(\omega)$, Eq. (S1) can be transformed into Eq. (S5):

$I(2\omega) = |-l + m \pm in|^2 I^2(\omega)$ (S5)

And the difference of SHG intensity upon the circular polarization stimulation is expressed as:

$\Delta I(2\omega) = I(2\omega)_{left} - I(2\omega)_{right} = 4Im((-l+m)n^*)I^2(\omega)$ (S6)

Hence, circular difference effect will occur if $(-l+m)$ and $n$ are simultaneously nonvanishing and if there is a phase difference between $(-l+m)$ and $n$.[20-24]

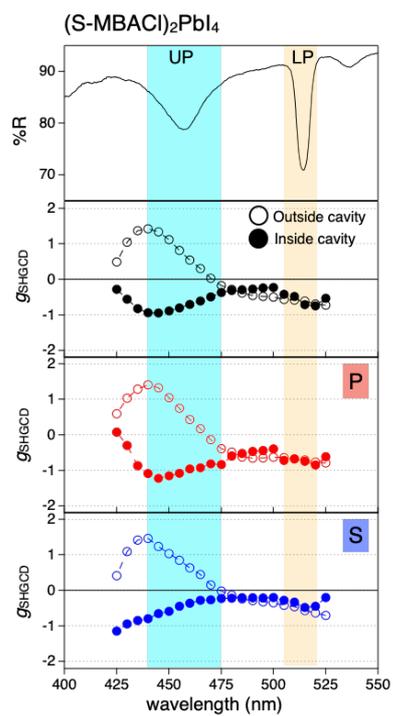

**Figure S12,** Polarization resolved g$_{SHG\text{-}CD}$ spectrum. In order from top to bottom, Rabi splitting reflectance spectrum at 45°, g$_{SHG\text{-}CD}$ spectrum without a polarizer, g$_{SHG\text{-}CD}$ spectrum in *p*-polarized condition and *s*-polarized condition.

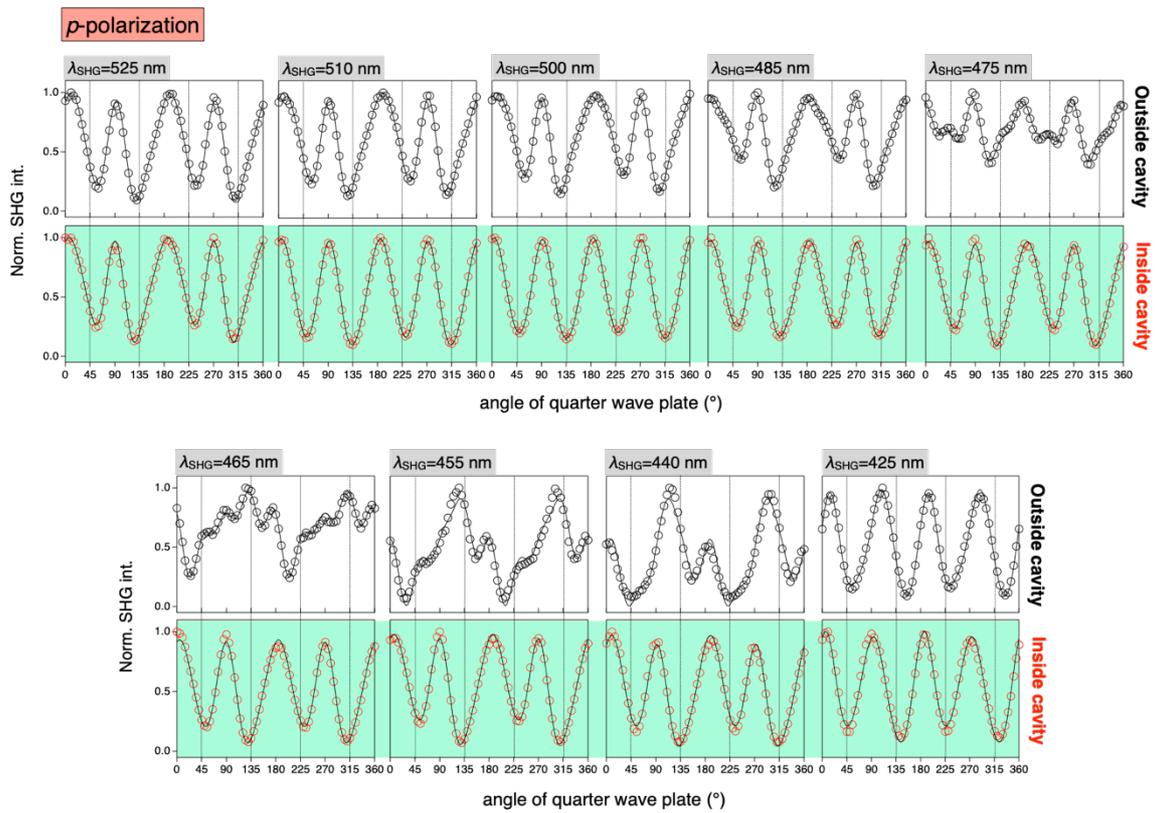

**Figure S13**, The normalized SHG intensity as function of QWR angle outside (Top) and inside (bottom) cavity in *p*-polarized conditons. The dots show experimental value and black solid line shows fitting curve.

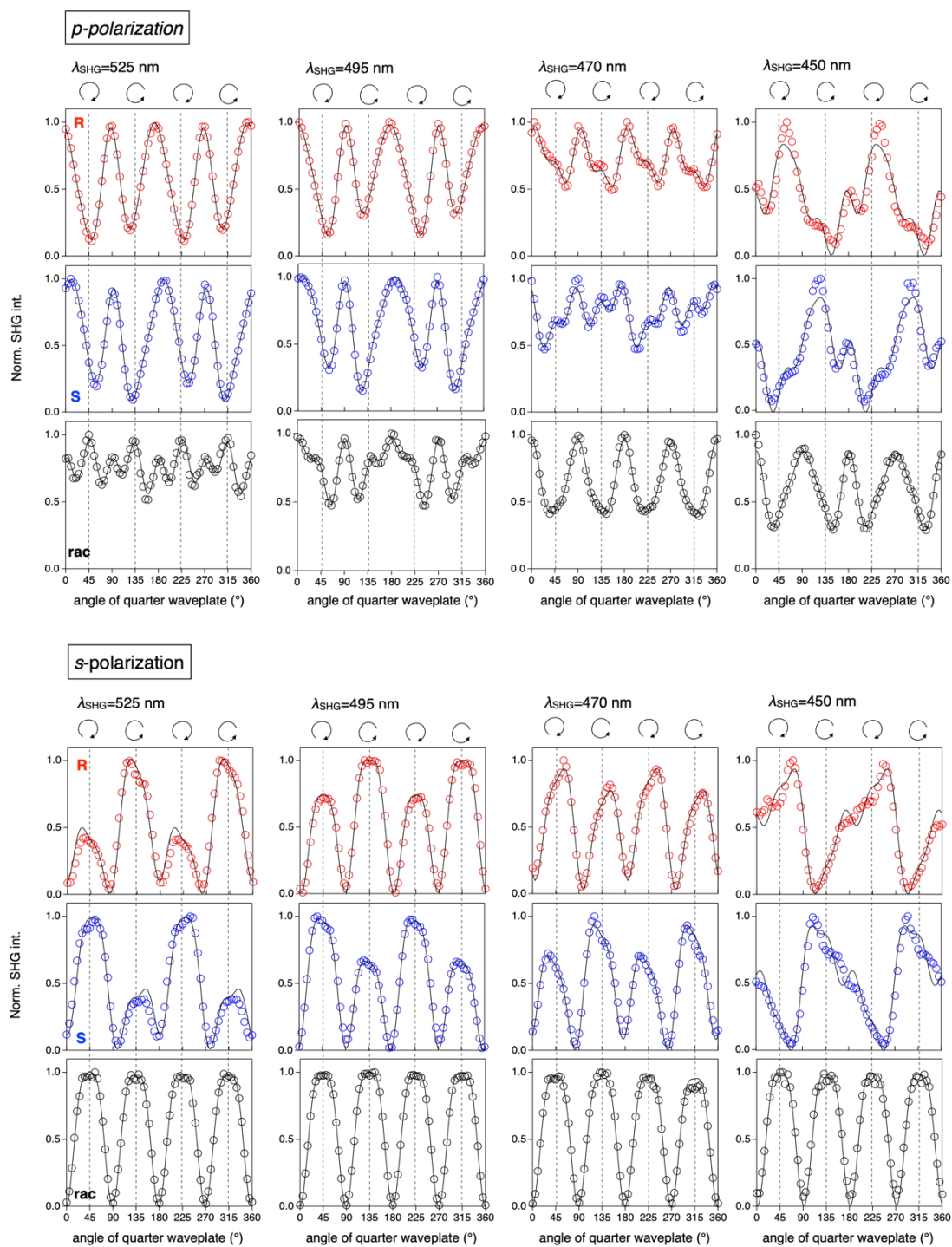

**Figure S14,** the normalized SHG intensity of R-/S-/racemic-(MBACl)$_2$PbI$_4$ thin film prepared on quartz substrate (outside-cavity) as function of QWR angle at several SHG wavelength in *p*- and *s*-polarized conditions. The dots and black solid line show experimental data and fitting curve, respectively.

| SHG(nm) | $l_p$(achiral) | $m_p$(achiral) | $n_p$ (chiral) | $l_s$ (chiral) | $m_s$ (chiral) | $n_s$ (achiral) |
|---|---|---|---|---|---|---|
| 525 nm | 1 | 0.03[±0.01]+i0.38[±0.01] (R)<br>-0.12[±0.01]+i0.52[±0.01] (S)<br>-1.16[±0.01]-i0.13[±0.02] (rac) | -(0.29[±0.01]+i0.06[±0.01])(R)<br>0.37[±0.01]+i0.07[±0.01](S)<br>0.10[±0.02]+i0.02[±0.01](rac) | 0.23[±0.02]+i0.14[±0.02] (R)<br>-(0.23[±0.02]+i0.14[±0.02]) (S)<br>0.04 [±0.01](rac) | -(0.18[±0.05]+i0.11[±0.02])(R)<br>0.25[±0.05]+i0.13[±0.03] (S)<br>-0.04 [±0.01](rac) | 1 |
| 495 nm | 1 | -0.28[±0.01]+i0.49[±0.01](R)<br>-0.44[±0.01]+i0.63[±0.01] (S)<br>-0.81[±0.01]+i0.41[±0.01] (rac) | -(0.03[±0.01]+i0.11[±0.01])(R)<br>0.14[±0.01]+i0.10[±0.01] (S)<br>0.06[±0.01]-i0.02[±0.004](rac) | -(0.06[±0.01]-i0.07[±0.01]) (R)<br>0.03[±0.01]-i0.07[±0.01] (S)<br>0.02[±0.003] (rac) | 0.06[±0.02]-i0.01[±0.01] (R)<br>0.02[±0.03]+i0.03[±0.01] (S)<br>-0.02 [±0.01] (rac) | 1 |
| 470 nm | 1 | -0.64[±0.01]+i0.03[±0.01](R)<br>-0.77[±0.01]-i0.06[±0.01] (S)<br>-0.33[±0.01]-i0.10[±0.01] (rac) | 0.40[±0.02]+i0.04[±0.01] (R)<br>-(0.36[±0.02]+i0.11[±0.01]) (S)<br>-0.02 [±0.01] (rac) | -(0.22[±0.03]+i0.06[±0.01]) (R)<br>0.21[±0.02]+i0.07[±0.02] (S)<br>0.03[±0.01] (rac) | 0.11[±0.02]+i0.03[±0.01] (R)<br>-(0.07[±0.02]+i0.01[±0.01]) (S)<br>-0.04[±0.02]+i0.01[±0.01](rac) | 1 |
| 450 nm | 1 | -1.02[±0.06]-i0.84[±0.08](R)<br>-1.02[±0.05]-i0.91[±0.06] (S)<br>-0.45[±0.01]-i0.45[±0.02] (rac) | -(0.78[±0.08]-i0.36[±0.05]) (R)<br>0.70[±0.07]-i0.32[±0.04] (S)<br>0.09[±0.02]+i0.02[±0.01] (rac) | -(0.62[±0.01]+i0.40[±0.01]) (R)<br>0.57[±0.07]+i0.37[±0.04] (S)<br>-0.06 [±0.01] (rac) | 0.44[±0.05]+i0.16[±0.02] (R)<br>-(0.45[±0.04]+i0.19[±0.02]) (S)<br>0.07 [±0.03] (rac) | 1 |

**Table S1**, Relative fitting values of profiles shown in Figure S14. Chiral coefficients of racemic-(MBACl)$_2$PbI$_4$ sample are almost negligible, while the R-/S-(MBACl)$_2$PbI$_4$ samples exhibit finite $n_p$, $l_s$, $m_s$ values with same magnitude and opposite signs to each other chirality.

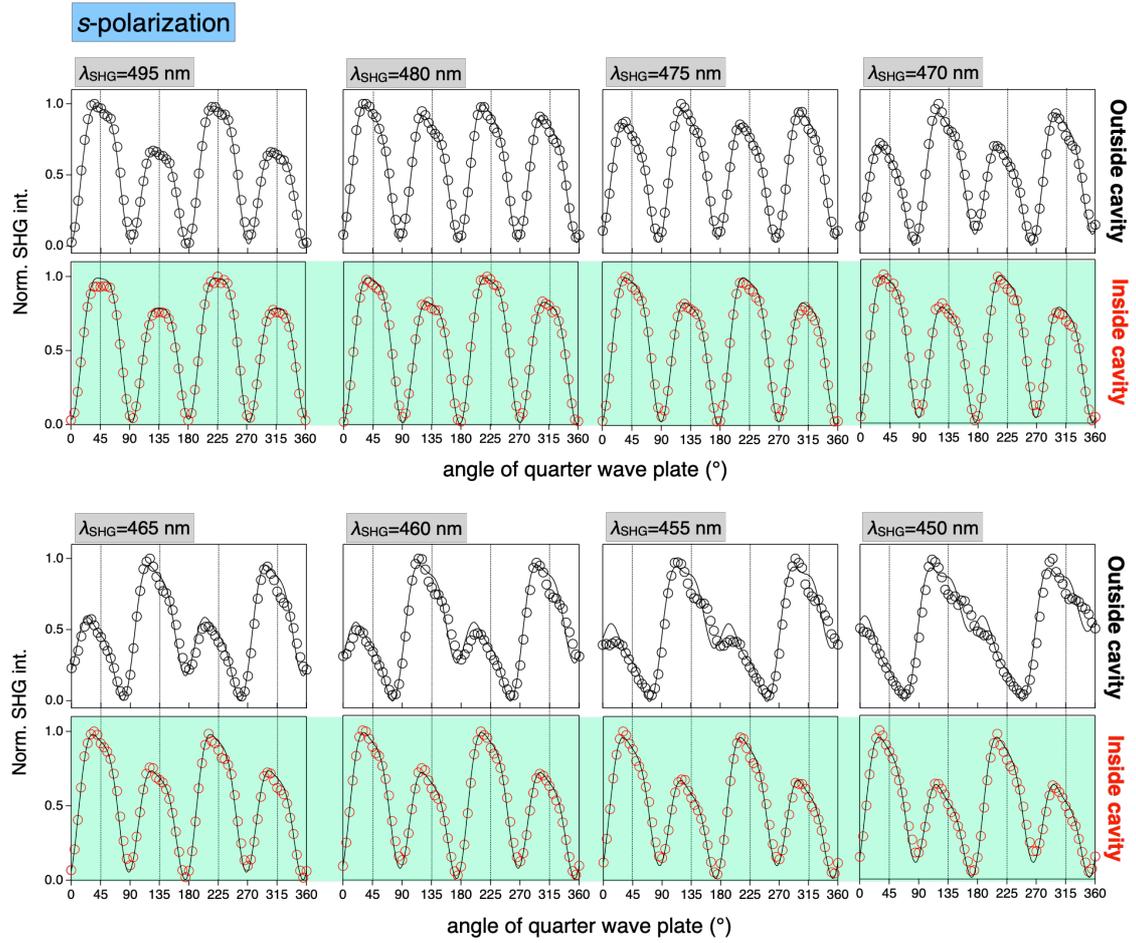

**Figure S15,** The normalized SHG intensity of outside- (Top) and inside-cavity (bottom) as function of QWR angle in *s*-polarized conditons. The dots show experimental value and black solid line shows fitting curve. Here, the data is shown as normalized SHG intensity. However, for the calculation of $|n_p + l_s|$, the fitting is conducted in actual SHG intensity data.